\documentclass[final,,authoryear,5p,times,twocolumn]{elsarticle}

\usepackage{graphicx}
\graphicspath{{./pdf/}}
\usepackage{url}
\usepackage{amssymb}

\journal{Radiation Measurements}

\begin{document}

\begin{frontmatter}

\title{Scintillation properties of pure and Ce$^{3+}$-doped SrF$_2$ crystals}

\author[igc,isu]{R. Shendrik\corref{cor1}}
\ead{shendrik@ieee.org}
\author[igc,isu]{E.~A.~Radzhabov}
\author[igc,isu]{A.~I.~Nepomnyashchikh}
\address[igc]{Vinogradov Institute of geochemistry SB RAS, Favorskogo 1a, Irkutsk,Russia, 664033}
\address[isu]{Physics department of Irkutsk state university, Gagarina blvd 20, Irkutsk, Russia, 664003}
\cortext[cor1]{Corresponding author: Vinogradov Institute of geochemistry SB RAS, Favorskogo 1a, Irkutsk,Russia, 664033}
\begin{abstract}
In this paper results of scintillation properties measurements of pure and Ce$^{3+}$-doped strontium fluoride crystals are presented. We measure light output, scintillation decay time profile and temperature stability of light output. X-ray excited luminescence outputs corrected for spectral response of monochromator and photomultiplier for pure SrF$_2$ and SrF$_2$-0.3 mol.\% Ce$^{3+}$ are approximately 95\% and 115\% of NaI-Tl emission output, respectively. A photopeak with a 10\% full width at half maximum is observed at approximately 84\% the light output of a NaI-Tl crystal after correction for spectral response of photomultiplier, when sample 10x10~mm of pure SrF$_2$ crystal is excited with 662 KeV photons. Corrected light output of SrF$_2$-0.3 mol.\% Ce$^{3+}$ under 662 KeV photon excitation is found at approximately 64\% the light output of the NaI-Tl crystal.
\end{abstract}


\end{frontmatter}

\section{Introduction}

The interest in new scintillation materials is promoted by increasing number of new applications in medicine, science, and homeland security, which require ramp-up of material production. The most perspective scintillators are bromides and iodides doped with Ce$^{3+}$ and Eu$^{2+}$ ions, such as SrI$_{2}$-Eu and LaBr$_{3}$-Ce. These crystals have high light outputs (up to 100000 photons/MeV for SrI$_{2}$-Eu), good energy resolution, and high proportionality \citep{Dorenbos10}. Disadvantages of these scintillators are high hygroscopic and price. In addition, SrI$_{2}$-Eu has temperature instability of light output observed by \citealp{Alekhin11}.

For the most applications a cheaper NaI-Tl scintillator has quite properties (light output about 45000 photons/MeV cited in \citealp{Berkley}). Therefore, one of the way in development of new scintillators is to find new materials with similar to NaI-Tl properties but no hygroscopic. In this way advanced materials for new scintillators are alkali-earth fluorides doped with rare earth ions. Theoretical limit of light output for them is up to 50000 photons/MeV \citep{Dorenbos10}. If an efficient energy transfer is provided then alkali-earth fluorides will be promising scintillators. A real light output of CaF$_2$-Eu is 18000-22000 photons/MeV, but BaF$_2$ and BaF$_2$-Ce crystals demonstrate lower light output at about 10000 photons/MeV \citep{Visser1991}. Scintillation properties of SrF$_2$ crystals are almost not investigated. Light output of SrF$_2$ was estimated about 10000-12000 photons/MeV by \citealp{Schotanus87}. However, potential light output of SrF$_2$ will be higher. Also SrF$_2$ crystals doped with Ce$^{3+}$ and Pr$^{3+}$ have a temperature stability of light output in wide range (20 $^\circ$C to 200 $^\circ$C) \citep{Shendrik2010}. Therefore, SrF$_2$ can be high-potential scintillator for well-logging. So, scintillator properties of strontium fluoride crystals are among the least studied of fluorides crystals, but these crystals have potential applications. Thus, the investigations of scintillation properties of strontium fluorides are topical today. This paper describes the scintillation properties of pure and cerium doped strontium fluorides crystals, a newly discovered inorganic scintillator.

 \section{Experimental methodology}

Growing with addition of CdF$_{2}$ as an oxygen scavenger, oxygen-free crystals of pure SrF$_2$ and doped with different concentrations of Ce$^{3+}$ ions were grown in a graphite crucible by the Stockbarger method.
We applied several experimental techniques in measurement of scintillation properties of the crystals. To determine light output, pulsed-height spectra under $^{137}$Cs 662 KeV gamma source excitation were measured with PMT FEU-39A, a homemade preamplifier and an Ortec 570A spectrometric amplifier. The crystals of 10x10 mm dimensions were polished and then covered with several layers of ultraviolet reflecting Teflon tape (PTFE tape). The shaping time of Ortec 570 spectrometric amplifier was set at 10 $\mu$s to collect much light from scintillator.

X-ray excited luminescence was performed using x-ray tube with Pd anode operating at 35 kV and 0.8 mA. The spectra were recorded at photon-counting regime using PMT FEU-39A and vacuum grating monochromator VM-4.

Scintillation decay time profiles under $^{137}$Cs E=662 KeV gamma source excitation were recorded by 200~MHz oscilloscope Rigol DS-1202CA. To register decay curves in wide time interval, we used oscilloscope input resistance set (50 $\Omega$ and 2.8 k$\Omega$).

 \section{Experimental results and discussion}

\begin{figure}[]
 \includegraphics[width=\columnwidth]{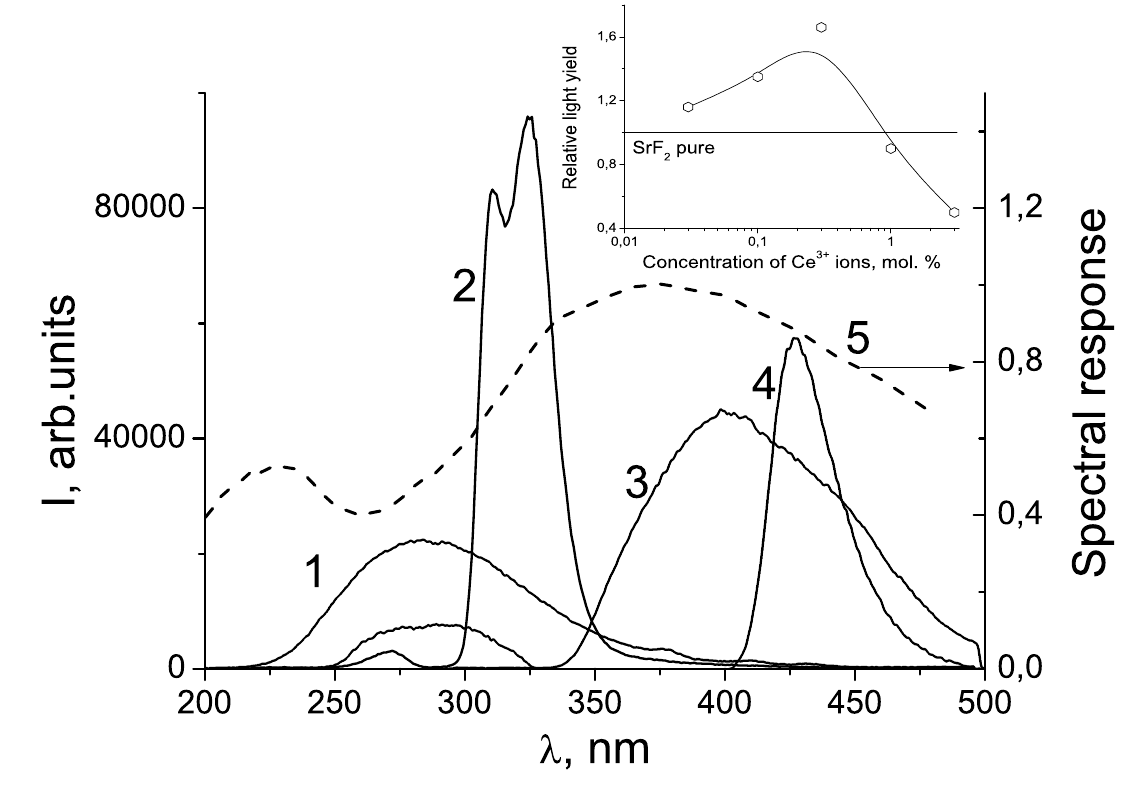}
 \caption{Luminescence spectra of SrF$_2$ (curve 1), SrF$_2$-0.3 mol.\% Ce$^{3+}$ (curve 2), NaI-Tl (curve 3), and CaF$_2$-0.1 mol.\% Eu$^{2+}$ (curve 4) under x-ray excitation. Dashed curve 5 is spectral sensitivity of the grating monochromator and PMT registration channel. In the inset dependence of integral intensity of emission bands of Ce$^{3+}$ ions versus Ce concentration is presented.}
 \label{spectra}
\end{figure}

Figure~\ref{spectra} shows spectra of x-ray luminescence of pure SrF$_2$, NaI-Tl, SrF$_2$-0.3 mol.\% Ce$^{3+}$, and CaF$_2$-0.1 mol.\% Eu$^{2+}$. In the spectrum of SrF$_2$ a wide band at 280 nm is attributed to self-trapped exciton (STE) emission. In SrF$_2$ doped with Ce$^{3+}$ ions STE luminescence is quenched and vanished at concentrations Ce$^{3+}$ ions higher than 1 mol.~\%. The most intense bands in x-ray luminescence spectra of SrF$_2$-Ce$^{3+}$ crystal at 310 and 325 nm correspond to 5d-4f emission of Ce$^{3+}$ ions. 

Luminescence spectrum of CaF$_2$-Eu$^{2+}$ is shown in Fig.~\ref{spectra}, curve 4. Its emission band is centered at 425 nm. This luminescence is due to 4f$^6$5d$^1$-4f$^7$ transitions in the Eu$^{2+}$ ion \citep{Kobayasi1980}.

Dependence of integral intensity of Ce$^{3+}$ ions emission bands on Ce concentration is shown in the inset to Figure~\ref{spectra}. The highest light output is demonstrated by SrF$_2$-0.3 mol.\% Ce$^{3+}$.

X-ray excited luminescence output measured by integral intensity is compared with the one of NaI-Tl crystal (Table~\ref{table1}). Light output of NaI-Tl crystals is approximately 43000 photons/MeV. Therefore, light output of the measured samples can be estimated. The data are shown in Table~\ref{table1}. Light output of CaF$_2$-0.1 mol.\% Eu$^{2+}$ is about 21500 photons/MeV that is in according with known data for CaF$_2$-Eu crystals given by \citealp{Berkley}. Pure SrF$_2$ has light output about 20640 photons/MeV, doped with 0.3 mol.\% and 1 mol.\% crystals have the ones about 34000 photons/MeV and 18500 photons/MeV, respectively.

\begin{table*}[!t]
\renewcommand{\arraystretch}{1.2}
\begin{minipage}{\textwidth}
\centering
\caption{Light output of SrF$_2$, SrF$_2$-Ce$^{3+}$, NaI-Tl and CaF$_2$-0.1 mol.\% Eu$^{2+}$ crystals measured under gamma and x-ray excitation.}
\label{table1}
\begin{tabular}{|c|c|c|c|c|}
\hline
\hline
Crystal & \multicolumn{2}{c|}{Light output of x-ray}  & \multicolumn{2}{c|}{Light output measured by }\\
 & \multicolumn{2}{c|}{excited luminescence spectra}  & \multicolumn{2}{c|}{pulse height spectra}\\
\hline
   & rel.units & photons/MeV & rel.units & photons/MeV \\
\hline
NaI-Tl & 1 & 43000 & 1 & 43000 \\
\hline
CaF$_2$-0.1 mol.\% Eu$^{2+}$ & 0.5 & 21500 & 0.44 & 18920 \\
\hline
SrF$_2$ & 0.48 & 20640 & 0.42 & 18060 \\
\hline
SrF$_2$-0.3 mol.\% Ce$^{3+}$ & 0.79 & 33970 & 0.32 & 13760 \\
\hline
SrF$_2$-1 mol.\% Ce$^{3+}$ & 0.43 & 18490 & 0.2 & 8600 \\
\hline
\hline
\end{tabular}
\end{minipage}
\end{table*}

All integral intensities and light outputs are presented without correction for spectral response of registration channel. The spectral response curve is shown in Figure~\ref{spectra}, dashed line. The sensitivity of the PMT and monochromator system in SrF$_2$ and SrF$_2$-Ce luminescence spectral range is lower than in NaI-Tl and CaF$_2$-Eu emission region. After the correction, light output of pure SrF$_2$ luminescence is about 40000 photons/MeV, and SrF$_2$ doped with 0.3 mol.~\% and 1 mol.~\% crystals have light outputs about 50000 photons/MeV and 28000 photons/MeV, respectively.

\begin{figure}[]
 \includegraphics[width=\columnwidth]{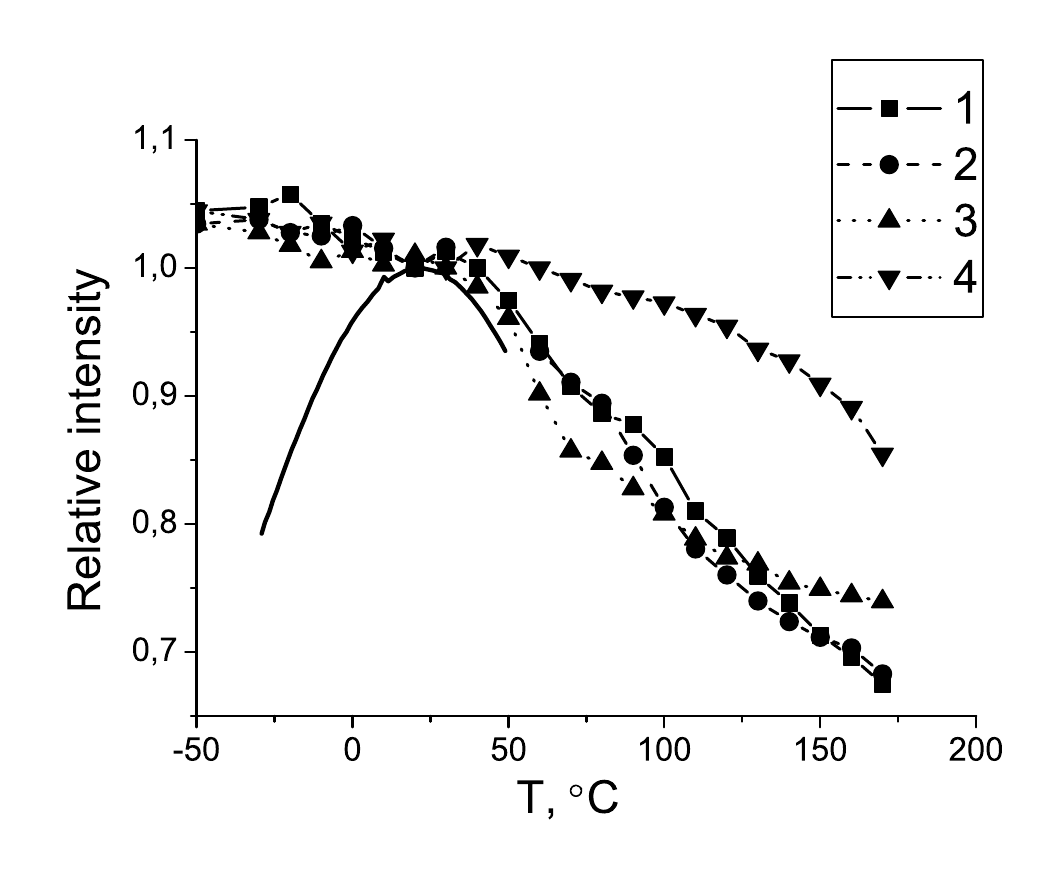}
 \caption{Temperature dependences of relative light output of SrF$_2$ crystals doped with 0.01 mol.~\% Ce$^{3+}$ (curve 1); 0.1 mol.~\% Ce$^{3+}$ (curve 2); 0.3 mol.~\% Ce$^{3+}$ (curve 3) and 1 mol.~\% Ce$^{3+}$ (curve 4) in comparison with NaI-Tl (solid line) measured by \citealp{Moszynski2006}.}
 \label{temp}
\end{figure}

Temperature dependences of x-ray excited luminescence intensity of SrF$_2$-Ce are given in Figure~\ref{temp}. Emission intensity does not depend on temperature in the range between -50 $^\circ$C and +50 $^\circ$C. At higher temperatures decrease of 5d-4f luminescence intensity is observed. At 170 $^\circ$C temperature light output of SrF$_2$ doped with 0.01 mol.\% and 0.1 mol\% of Ce$^{3+}$ ions decreases to 30~\%. 25 \% and 15 \% decreases in integral emission are found in the SrF$_2$ crystals doped with 0.3 mol.~\% Ce$^{3+}$ and 1 mol.~\%, respectively. Crystals of SrF$_2$-0.3 mol.~\% Ce$^{3+}$ and SrF$_2$-1 mol.~\% Ce$^{3+}$ demonstrate high temperature stability of light output in the region between -50 $^\circ$C and 170 $^\circ$C in comparison with NaI-Tl (see Fig.~\ref{temp}, solid line). For this reason, the SrF$_2$-Ce$^{3+}$ crystals would be perspective scintillators for well-logging applications.

\begin{figure}[]
 \includegraphics[width=\columnwidth]{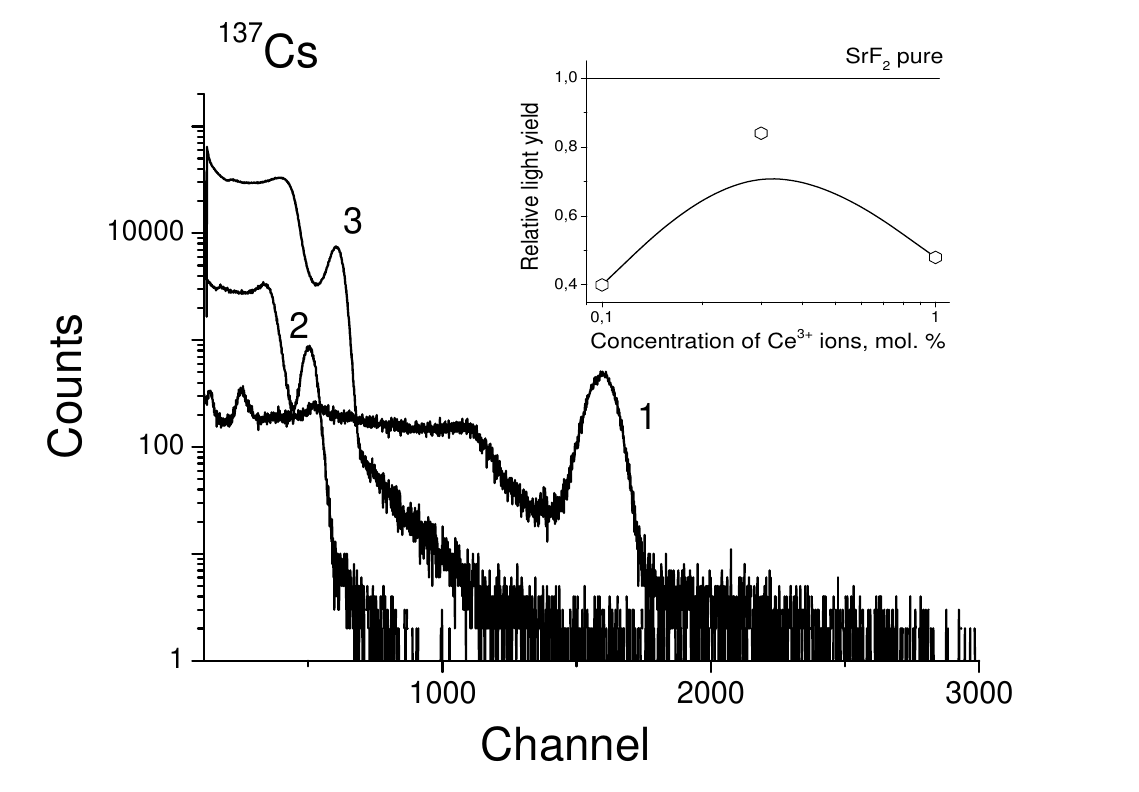}
 \caption{Pulse height spectra of NaI-Tl (curve 1), SrF$_2$-0,3 mol.\% Ce$^{3+}$ (curve 2) and SrF$_2$ (curve 3) under gamma-source $^{137}$Cs (E=662 KeV) excitation. In the inset dependence of relative light outputs of SrF$_{2}$-Ce crystals versus Ce concentration is presented.}
 \label{pulsed}
\end{figure}

Figure~\ref{pulsed} shows pulse height spectra of SrF$_2$, SrF$_2$-0.3 mol.~\% Ce$^{3+}$ and NaI-Tl. The photopeak corresponding to the $^{137}$Cs energy photon is seen in each curve in the Figure~\ref{pulsed}. Light output of SrF$_2$ and SrF$_2$-Ce crystals was measured by comparing these response to 662 KeV energy to the response of NaI-Tl crystal with known characteristics under the same conditions. The photopeak in pure SrF$_2$ is centered at a pulse height that is 42\% of the 662~keV photopeak pulse height in NaI-Tl. Using the NaI-Tl light output of 43000 photons/MeV, this implies that the light output of pure SrF$_2$ is approximately 18000 photons/MeV that is similar to x-ray emission light output (see Table \ref{table1}). The full width at half maximum (FWHM) in pure SrF$_2$ of the 662 keV photopeak is 10~\%. FWHM of NaI-Tl is 6.7~\%, which is consistent with known results given by scintillators database of \citealp{Berkley}.

Dependence of SrF$_2$-Ce relative light output versus concentration is demonstrated in the inset to Figure~\ref{pulsed}. Best light output of the Ce-doped crystals is found for SrF$_2$-0.3 mol.~\% Ce$^{3+}$. Its photopeak is centered at a pulse height that is 32~\% of the 662~keV photopeak pulse height in NaI-Tl, this implies that the light output of SrF$_2$-0.3 mol.\% Ce$^{3+}$ is approximately 14000 photons/MeV. FWHM of the SrF$_2$-0.3 mol.\% Ce$^{3+}$ is approximately 9.3~\%, which is lower than FWHM of pure SrF$_2$. This fact means that in spite of worse light output SrF$_2$-0.3 mol.\% Ce$^{3+}$ crystal has better energy resolution in comparison with pure strontium fluoride.

All light outputs without any corrections are given in Table~\ref{table1}. Bearing in mind that spectral sensitivity of S20 photocathode (PMT FEU 39A) is higher at CaF$_{2}$-Eu and NaI-Tl emission bands than in SrF$_{2}$ and SrF$_{2}$-Ce luminescence region \citep{Photonis}. Corrected light outputs of pure SrF$_2$ is about 80\% (36000~photons/MeV) of NaI-Tl and SrF$_2$-0.3 mol.~\% of Ce$^{3+}$ -- about 60\% (26000~photons/MeV) of NaI-Tl.

\begin{figure}[]
 \includegraphics[width=\columnwidth]{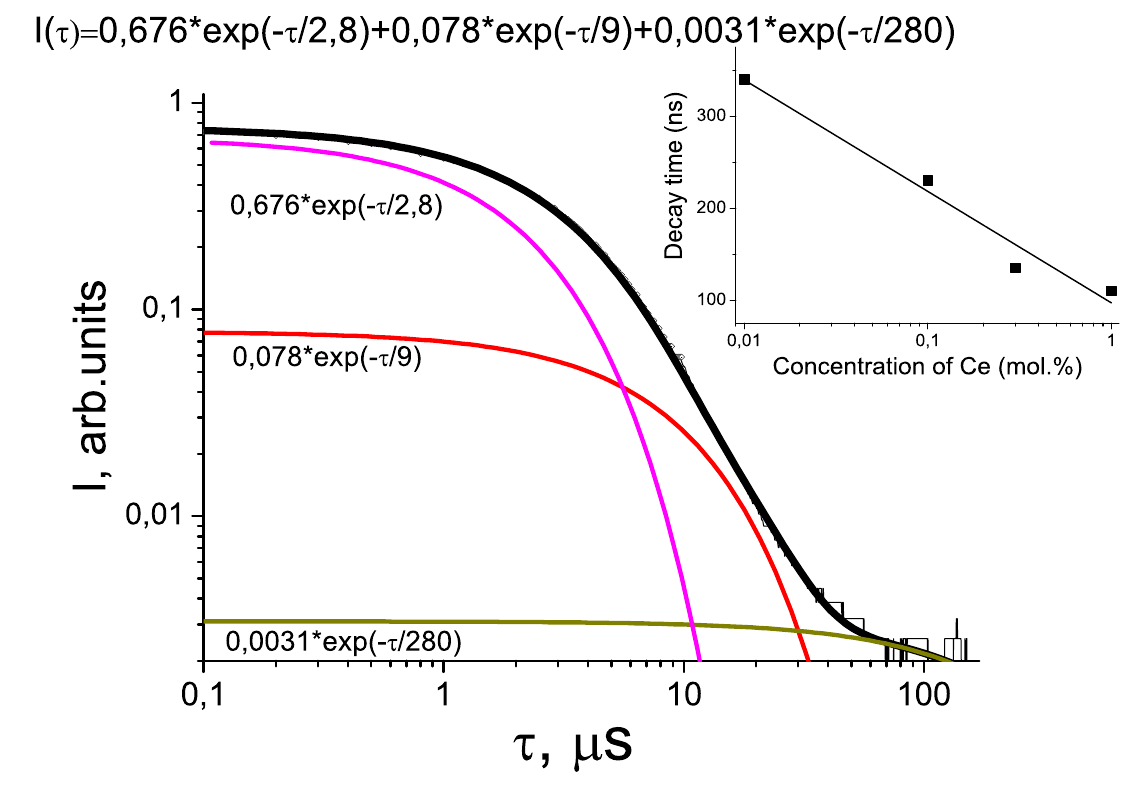}
 \caption{Scintillation decay time profile of SrF$_2$-0.3 Ce$^{3+}$ crystal measured under gamma-source $^{137}$Cs excitation (E=662 KeV). Exponential components of total decay curve are presented separately. In the inset dependence of shortest decay time component on Ce$^{3+}$ ions concentration is shown.}
 \label{decay}
\end{figure}

Scintillation decay time profile of SrF$_2$-0.3 mol.\% Ce$^{3+}$ is shown in Figure~\ref{decay}. Resistance of oscilloscope input was 2.6 K$\Omega$ for registration long time decay components in Ce$^{3+}$ emission. The decay time is described by a sum of exponents. First component (2.8 $\mu$s) in Ce$^{3+}$ decay is integrated short components. Lifetime of the shortest one equals 130~ns at 50~$\Omega$ input resistance in SrF$_2$-0.3 mol.\% Ce$^{3+}$ and it becomes longer with decrease of Ce concentration. Decay constants of this component in dependence on Ce concentration are presented in the inset of the Figure~\ref{decay}. 

Contribution of slow components to scintillation time profile is estimated. In the figure~\ref{decay} exponential components of total decay curve are shown separately. There are two long time components in SrF$_2$-0.3 mol.\% Ce$^{3+}$ emission. 20\% of the light is emitted with a 9~$\mu$s time constant, and 25\% of the light is emitted with a 280~$\mu$s time constant.

Emission decay time of cerium doped alkaline-earth fluorides under optical excitation at lowest energy absorption bands is estimated about 30 ns \citep{Visser93, Radzhabov2004, Wojtowicz00-2}. Under vacuum ultraviolet excitation at exciton and higher energies regions the decay of Ce-doped fluorides became nonexponential \citep{Wojtowicz00-2}. Under x-ray excitation the decay curves is also~nonexponential~(Fig.~\ref{decay}). 

Whole decay curve can be described by several processes. Fast stage could be ascribed to resonance energy transition in nearest pairs of exciton and cerium ion. In the inset of the figure~\ref{decay} concentration dependence of these decay constants is presented. Note that the decay becomes shorter with increasing Ce$^{3+}$ ions concentration due to reduction of distance between exciton and activator ion with increase of cerium ions concentration. 

In SrF$_{2}$-Ce crystals thermoluminescence glow peaks at 200-250~K were found \citep{Radzhabov01,Maghrabi01}. Broad glow peaks at higher temperatures are shifted to lower temperature with increasing concentration of Ce$^{3+}$ ions in the SrF$_{2}$ crystal \citep{Maghrabi01}. Long stages in scintillation time profile can be attributed to thermoactivated processes related to electron or hole delayed transfer to the activator ion. A similar energy transfer mechanism has been observed in SrF$_{2}$ doped with Pr \citep{Shendrik2012}.

Therefore, the difference in light outputs measured from x-ray luminescence and pulse height spectra is explained by presence of intensive slow components in cerium ions luminescence (see Fig.~\ref{decay}). They give a large contribution to total light output. Shaping time of pulse height spectrum measurement is 10~$\mu$s, and a large part of emitted light is not registered whereas a rate of x-ray excited luminescence spectra registration is amount about 1~s$^{-1}$ that leads to much more light registration. 

Light output of SrF$_2$-Ce$^{3+}$ samples can be increased by decreasing of slow component contribution in Ce$^{3+}$ luminescence. It might be possible by co-doping these crystals with Ga$^{3+}$, In$^{3+}$ or Cd$^{2+}$ ions to change band gap in SrF$_{2}$-Ce crystal and, thence, reducing the role of traps in scintillation energy transfer. This idea has been made in garnets by \citealp{Fasoli11}. However, Cd$^{2+}$ ions bring to STE suppressing in alkali-earth fluorides \citep{Radzhabov2005-jp, Radzhabov2005-pss}. Consequently, Cd$^{2+}$ co-doping leads to suppression of efficient resonance exciton energy transfer mechanism and following light output reducing. Therefore, Cd$^{2+}$ co-doping is not eligible way for increasing light output of SrF$_{2}$-Ce scintillator. A role of Ga$^{3+}$ and In$^{3+}$ impurities in STE suppressing has not yet been investigated and follow-up study of the crystals doped with Ga$^{3+}$ and In$^{3+}$ ions is required.

\section{Conclusion}

The SrF$_2$ crystal are well suited for use as gamma radiation detector. It has a higher (4.18 g/cm$^{3}$) than NaI-Tl (3.67 g/cm$^{3}$) density, comparable light output, and it is no hygroscopic. 
Taking into account that crystals SrF$_{2}$ - 0.3 mol.\%~Ce$^{3+}$ and SrF$_{2}$ - 1 mol.\%~Ce$^{3+}$ have a high temperature stability of light output in the temperature interval from -50~$^\circ$C to 200~$^\circ$C, these materials can be applied in well-logging scintillation detectors. Summarizing the experimental results we conclude that strontium fluoride crystals would be useful as newly perspective scintillator.

\section*{Acknowledgement}

This work was partially supported by Federal Target Program "Scientific and scientific-pedagogical personnel of innovative Russia" in 2009-2013 and Russian Foundation for Basic Research (RFBR).

Authors are thankful to V. Kozlovskii for growing the crystals investigated in this work.

\bibliographystyle{elsarticle-harv}
\bibliography{Biblio}

\end{document}